\pdfoutput=1
\documentclass[11pt,letterpaper]{article}
\usepackage[T1]{fontenc}
\usepackage{lmodern}
\usepackage[utf8]{inputenc}
\usepackage{amsmath,amssymb}
\usepackage[margin=1.05in]{geometry}
\usepackage{microtype}
\usepackage{xcolor}
\definecolor{paperlink}{RGB}{0,0,114}
\usepackage{graphicx}
\usepackage{booktabs,tabularx,array,multirow}
\usepackage[font=small,labelfont=bf]{caption}
\usepackage[round,authoryear]{natbib}

\usepackage{placeins}

\usepackage[hyphens]{url}
\usepackage{hyperref}
\hypersetup{colorlinks=true,linktoc=all,
  linkcolor=paperlink,citecolor=paperlink,urlcolor=paperlink,pdfstartview=Fit,
  pdftitle={Use-Case-Driven Ontology Modeling with the NeOn Methodology for Ambient Assisted Living},
  pdfauthor={Daniele Spoladore, Marco Sacco, Alberto Trombetta, Edoardo Battisti}}
\usepackage{enumitem}
\setlist[itemize]{leftmargin=2em,itemsep=0.25em,topsep=0.35em}
\setlist[enumerate]{leftmargin=2.25em,itemsep=0.25em,topsep=0.35em}
\usepackage{fancyhdr}
\date{}

\newcommand{\ont}[1]{\mbox{\textsf{#1}}}
\newcommand{\vo}{VO\textsubscript{2}max}
\newcommand{\chk}[1]{\textcolor{red}{#1}}

\newcounter{swrlrule}
\renewcommand{\theswrlrule}{R\arabic{swrlrule}}
\newenvironment{swrl}[1]{\par\medskip\refstepcounter{swrlrule}\label{#1}\noindent
  \begin{minipage}[c]{\dimexpr\linewidth-3.5em\relax}\raggedright\small\ttfamily
  \hangindent=1.5em\hangafter=1}
  {\end{minipage}\hfill(\theswrlrule)\par\medskip}
\newcommand{\rref}[1]{(\ref{#1})}
\newcommand{\imp}{$\;\rightarrow\;$}

\title{Use-Case-Driven Ontology Modeling with the NeOn Methodology\\for Ambient Assisted Living}
\author{Daniele Spoladore\thanks{Institute of Industrial Technologies and Automation (ITIA), National Research Council of Italy (CNR), Milan, Italy. Corresponding author: \href{mailto:daniele.spoladore@itia.cnr.it}{\texttt{daniele.spoladore@itia.cnr.it}}.}
\and Marco Sacco\footnotemark[1]
\and Alberto Trombetta\thanks{University of Insubria, Italy.}
\and Edoardo Battisti Sironi\thanks{Politecnico di Milano, Italy.}}

\begin{document}
\maketitle
\thispagestyle{plain}

\begin{abstract}
Research on smart homes and ambient assisted living (AAL) has grown steadily over the last decade, driven largely by the ageing of the population. A smart home has to coordinate heterogeneous smart objects in order to offer services tailored to its dwellers, and this requires a shared formal description of both the dwellers and the devices. We present the ontology developed as the semantic data layer of the Virtual Home Framework in the Italian Design For All project. Its requirements are derived from two use cases: a frail elderly woman who exercises at home on a cycle-ergometer, and a designer who configures a kitchen for users with different impairments. The requirements are collected in an Ontology Requirements Specification Document and the ontology is built with the NeOn methodology, reusing and reengineering existing ontologies where possible. The resulting modular ontology describes users through the International Classification of Functioning, Disability and Health (ICF) and through their cardiorespiratory fitness; it also describes smart objects, their programs, the measurements they produce, and the comfort dimensions of the home. SWRL rules classify fitness levels, suggest exercise loads, assess indoor comfort, and identify the smart objects suited to a given impairment.
\end{abstract}

\noindent\textbf{Keywords:} ontology development methodology; modular ontology; use case; ambient assisted living ontology.

\section{Introduction}\label{sec:intro}

Research on the smart home, ambient assisted living (AAL), and domestic healthcare has increased considerably over the last ten years. One reason is demographic: people aged 65 or over are expected to reach 30\% of the European population~\citep{giannakouris2008}. A smart home is expected to provide its inhabitants with services that help them live better, healthier, and safer lives in their own home. To do so, it needs a set of smart objects that cooperate to acquire, manage, communicate, and share knowledge about the domestic environment and its inhabitants. Integrating and managing such objects is not a trivial task~\citep{jiang2004}, and the health issues associated with ageing further require a consistent, shared conceptualization of the dwellers.

The Italian Design For All (D4All) project addresses these issues with semantic web technologies. Its aim is a sustainable smart home paradigm designed around real users: frail elderly people, people with impairments, and people without impairments. The core of the approach is an ontology that represents the dwellers together with the smart objects operating in the house. This paper describes how that ontology was developed and what it contains. The development starts from use cases typical of the smart home, follows the NeOn methodology~\citep{suarez2015}, and reuses existing ontologies where they fit.

The paper makes three contributions. First, it shows how personae and use cases can be turned into the requirements of an Ontology Requirements Specification Document (ORSD). Second, it describes a modular ontology that represents the user's health condition with the ICF and classifies cardiorespiratory fitness with SWRL rules. Third, it describes the modules for smart objects, measurements, comfort dimensions, and the configuration of living environments.

Section~\ref{sec:d4all} presents the Virtual Home Framework of the D4All project. Section~\ref{sec:related} reviews semantic models for AAL, both for interoperability and for the modeling of users and their activities. Section~\ref{sec:usecases} describes how the personae and the two use cases were built, and Section~\ref{sec:orsd} derives the ORSD from them. Section~\ref{sec:neon} explains which NeOn scenarios were applied, and Sections~\ref{sec:health} and~\ref{sec:smart} describe the two resulting ontologies. Section~\ref{sec:conclusions} summarizes the results and outlines future work.

\section{The Design For All project and the Virtual Home Framework}\label{sec:d4all}

Traditional design is oriented toward a standard individual, an abstraction of real men and women, and ignores many variables of the actual users, such as their skills and knowledge~\citep{mace1991}. Design for All, also known as Universal Design, instead takes the differences among users into account and proposes solutions that adapt to the needs of specific categories of users. Its principles apply naturally to AAL~\citep{costa2009}, where services are tailored to users in the specific context of the house.

The D4All project\footnote{\url{http://www.d4all.eu/en}} applies the guidelines of Universal Design to domotic and inclusive domestic environments. Designing such environments for people without impairments, elderly people, people with disabilities, or families requires heterogeneous tools over the whole lifecycle of the home, from design to use, control, and monitoring. These tools must account for the specific users and for the way they interact with the available functions and technologies. The interactions also generate heterogeneous data, which must be collected from the home and made available to the dwellers and to remote users such as caregivers or clinicians. The project therefore aims to provide different categories of users with tools that compensate for the limits of their physiological status and let them perform daily activities that would otherwise be precluded or strenuous. This is the Smart Home paradigm~\citep{harper2006}: a residence equipped with smart objects (appliances, sensors) that anticipate and respond to the needs of the inhabitants. Unlike in a traditional home, these objects are expected to work together reliably and predictably, sharing knowledge about the dwellers and their actions~\citep{koskela2004} in order to support their comfort and well-being.

Two research problems follow. The first is the description of users in terms of their skills, capabilities, limitations, and wishes, which calls for a shared paradigm able to describe different kinds of users. The second is semantic interoperability among the smart objects (sensors, software, domestic appliances), which calls for a formalization of the information produced in the home so that software tools can be integrated.

\begin{figure}[tbp]
\centering
\includegraphics[width=0.42\textwidth]{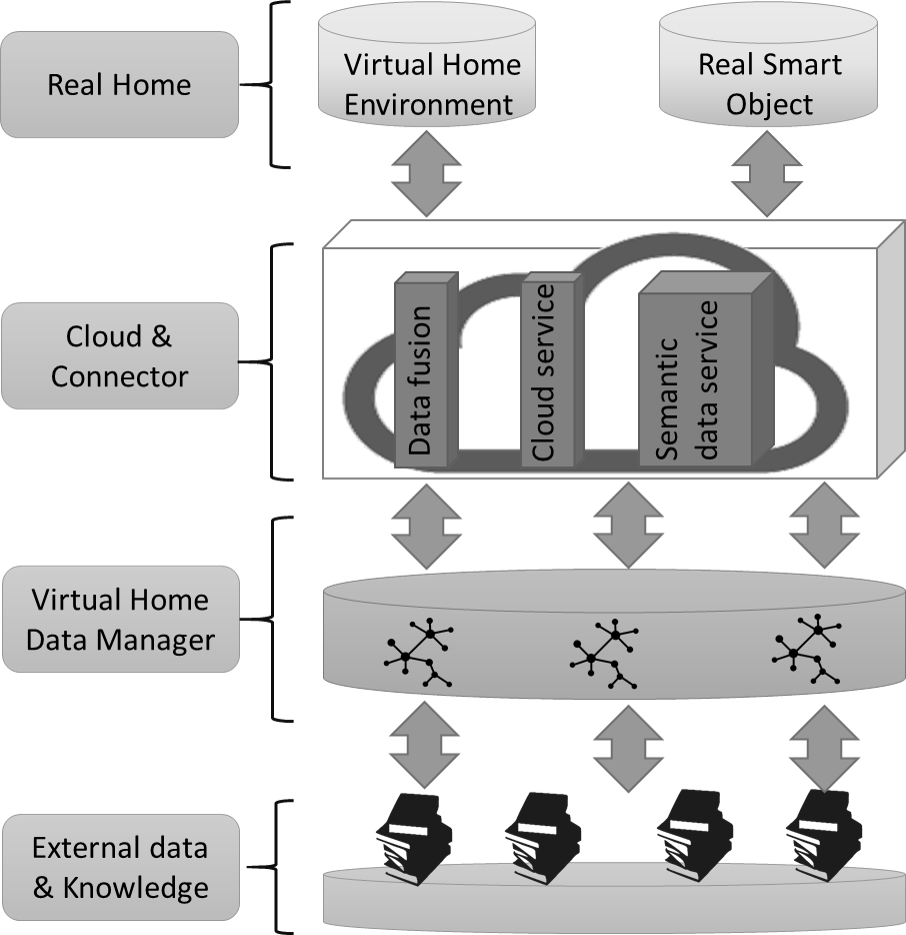}
\caption{The Virtual Home Framework.}\label{fig:vhf}
\end{figure}

The Virtual Home Framework (VHF) was conceived to address both problems. It is an integrated platform that manages the knowledge produced in the domestic environment and lets software tools exchange it~\citep{sacco2014}. As shown in Figure~\ref{fig:vhf}, it rests on three pillars:
\begin{itemize}
\item a semantic data model, the Virtual Home Data Manager (VHDM), which is the subject of this paper;
\item a cloud-hosted connection platform, described by~\citet{modoni2017};
\item a virtual representation of the home environment and of the real smart objects.
\end{itemize}
The VHDM gives a consistent and shared representation of the knowledge about the domestic environment, the users, and the smart objects. It must be extensible and have adequate granularity; it must support data consistency, safety, and management; and it must allow automated reasoning that adds derived knowledge about the concepts and their relationships.

\section{Related work}\label{sec:related}

Semantic interoperability among the devices of a smart home is one of the most discussed issues in AAL. \citet{welge2014} address the interoperability of devices and distributed systems with semantic web technologies, which they regard as a promising answer to problems of runtime behavior, heterogeneous structures, and component diversity; the outcome is a prototype ontology engineering process that collects and aggregates AAL domain knowledge. \citet{meshkova2008} present an ontology-based knowledge base of the domestic environment, describing the services offered to the dwellers and the hardware, software, and network components; developed in the European COMANCHE project, it is a scalable ontology for the conceptualization and management of home services. \citet{grguric2015} describe an upper ontology that captures the semantics of the AAL domain and the services of an AAL platform, including an orchestration module for data from sensors and other services. The W3C Semantic Sensor Network Incubator Group developed the Semantic Sensor Network (SSN) ontology~\citep{compton2012}, the result of an analysis of the most relevant sensor ontologies. SSN describes and classifies sensors by their capabilities and gives a process-oriented view of measurements, observations, and deployments.

The semantic modeling of dwellers and their activities has also received attention. \citet{ni2016} present a model of human activity in the smart home that infers activity-related information by reasoning over raw data and context. \citet{mocholi2010} describe an ontology of AAL services covering environments, contexts, users, and the interactions between users and environments, with attention to accessibility, habits, and abilities. \citet{razmerita2003} propose OntobUM, an ontology-based user modeling architecture in which the user supplies identity, contacts, preferences, and competencies, while intelligent services update the profile from the user's behavior. \citet{skillen2012} propose an ontology for personalizing context-aware applications that covers static and dynamic aspects of the user, including health, interests, preferences, and abilities. \citet{golemati2007} model static user information (physical characteristics, living conditions, profession, interests, education) together with the user's experience.

\section{Use cases}\label{sec:usecases}

The VHF involves a wide range of technologies, so the requirements were collected through use cases in which users perform everyday activities in their own house with the assistance of the framework. Use cases expose many aspects of an interactive system at design time: they make it possible to assess the impact of each design decision and to check whether general solutions exist for a specific issue~\citep{alexander2005}. The use cases were built with scenario-based design~\citep{cooper2004} in two phases, the identification of personae and the definition of use cases, and were then used to identify the domains the ontology must cover.

\subsection{Personae}\label{sec:personae}

The ontology must describe what users can and cannot do. This requires a balance between a medical and a sociological view, which accounts for both the physiological status of the users and the context in which they live. The first phase therefore identified personae, fictional characters that represent the types of users of the VHF~\citep{cooper2004}; personae also ease communication among the stakeholders of a project~\citep{pruitt2003}. Six personae were drafted with the clinical partners of the project:
\begin{enumerate}
\item \emph{Frail elderly user}: aged 65 or more, with an unstable health condition and at least one of the following: involuntary weight loss of 4.5\,kg in the last year; asthenia or fatigue on at least three days a week; limited physical activity and a sedentary lifestyle; low walking speed, namely more than 6\,s to walk 4\,m.
\item \emph{User with visual impairment}: a visual impairment that totally or partially compromises social and working activity.
\item \emph{User with hearing impairment}: partial or total hearing loss.
\item \emph{User with cognitive impairment}: alterations or dysfunctions of fundamental cognitive functions such as memory, perception, reasoning, and language; severe mental illnesses are excluded.
\item \emph{User with motor impairment}: multiple or specific neuro-motor deficits, including all pathologies of the neuro-motor system.
\item \emph{User without impairment}: none of the impairments above; this broad persona includes elderly people without specific conditions.
\end{enumerate}
Each persona covers a wide range of specific cases that must be explored further.

\subsection{Use cases}\label{sec:uc}

In the second phase the personae were placed in use cases. Following~\citet{benyon2005}, a use case is a type of scenario that describes the interaction between users and devices. Designing it means allocating tasks and functions to the actors, here both people and smart objects, and narrating their interactions. One use case was developed for each persona. We report two: the use case of the frail elderly persona, as an example of a user supported in a daily activity, and a use case in which a designer configures a smart kitchen, which involves several personae and exercises the classification capabilities of the ontology. The information gathered from the use cases was essential to compile the ORSD of Section~\ref{sec:orsd}, which follows the template of~\citet{suarez2009}.

\paragraph{Use case 1: an elderly woman exercising at home.}\label{sec:uc1}
Appropriate physical exercise helps elderly people prevent many health problems~\citep{nieman1993} as well as cognitive~\citep{laurin2001,geffken2001} and motor~\citep{grimby1986} impairments. The first use case describes a frail woman aged 72 (Simona in Table~\ref{tab:cq}) who must exercise daily on a cycle-ergometer in her bedroom. Because of her health condition she must be monitored during the exercise, and the data of each session must be available to caregivers and clinicians, who periodically assess improvements or deteriorations. Her home is equipped with the cycle-ergometer, a heart rate monitor, a breath rate monitor, and a blood pressure monitor. A tablet running the VHF gives her instructions, shows her performance in real time, and issues alerts based on the sensor readings. When she enters the bedroom a presence sensor detects her. She then ventilates the room and waits for it to reach 20\,\textdegree C. Once air quality and temperature are adequate, she wears the sensors and starts the exercise, whose load is set according to her health condition. Her physiological status is monitored throughout, and at the end the session data and physiological measurements are stored for the caregivers.

This simple use case raises three requirements. The user's health condition must be assessed both qualitatively and quantitatively; for this the project adopted the International Classification of Functioning, Disability and Health (ICF),\footnote{\url{http://www.who.int/classifications/icf/en/}} a unified, standard language for health-related components. The exercise load is based on cardiorespiratory fitness~\citep{farooqui2012}, which measures how well the circulatory and respiratory systems supply the muscles during activity and is associated with several health outcomes~\citep{kodama2009,sui2007}. Finally, data must flow among the sensors, the user, and the room, whose temperature and air quality condition the exercise.

\paragraph{Use case 2: configuring a kitchen for impaired users.}\label{sec:uc2}
In the second use case a designer uses an application to choose the most suitable appliances for users with mild or moderate impairments (personae~2 to~5). The designer acquires the blueprints of the user's kitchen, builds a virtual model, and places in it smart objects chosen from a list: oven, refrigerator, cooktop, dishwasher, and sensors for air quality, presence, and luminance. The project can then be saved, and other projects can be opened and modified. Preparing a meal is one of the Instrumental Activities of Daily Living (IADLs)~\citep{spector1987} that allow an individual to live independently. Offering each user the most appropriate solution requires a taxonomy of smart objects, a description of each appliance, and rules that match smart objects to the users for whom they are recommended. As in the first use case, the matching depends on a reliable classification of the user's health condition.

\section{Ontology requirements specification}\label{sec:orsd}

The use cases identify the key points the ontology must cover. Following a bottom-up approach, from the most concrete to the most abstract~\citep{fernandez2002}, these key points yield an initial set of concepts, which the NeOn methodology~\citep{suarez2015} and the ORSD refine. Starting from use cases simplifies the identification of the knowledge to represent and indicates the expected granularity and extent of the model. The ORSD also identifies the reusable ontological resources from which the whole model is then developed.

\begin{figure}[htbp]
\centering
\includegraphics[width=0.95\textwidth]{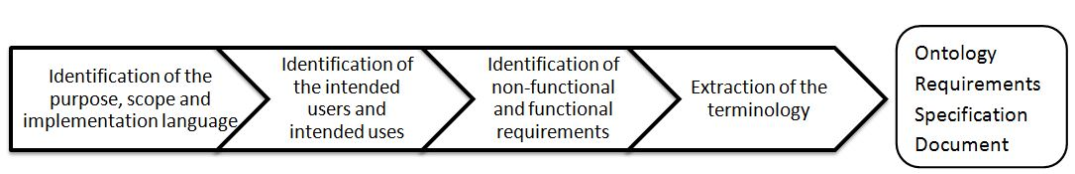}
\caption{Tasks for writing the ORSD (adapted from~\citealp{suarez2009}).}\label{fig:orsd}
\end{figure}

An ORSD states what an ontology supports, its domain of application, and its intended users and uses. It helps ontology engineers decide which concepts to include and which existing knowledge sources to reuse~\citep{sure2009}. The ORSD of the VHDM was compiled with the tasks defined by~\citet{suarez2009}, summarized in Figure~\ref{fig:orsd}.

\paragraph{Purpose, scope, and implementation language.}
The purpose of the ontology is to describe different kinds of users in a language understandable by clinicians, caregivers, and rehabilitation therapists, and to describe the smart objects in the user's home. Its scope comprises the users' health condition and cardiorespiratory fitness, the smart objects and the measurements they perform, and the classification of smart objects as suitable for specific types of users. The implementation languages are RDF\footnote{\url{https://www.w3.org/RDF/}} and OWL,\footnote{\url{https://www.w3.org/OWL/}} with SWRL\footnote{\url{https://www.w3.org/Submission/SWRL/}} rules; the ontology was developed with the open-source editor Prot\'eg\'e.\footnote{\url{http://protege.stanford.edu/}}

\paragraph{Intended end-users and uses.}
Clinicians, caregivers, and rehabilitation therapists access the users' health condition and cardiorespiratory fitness. Designers who configure smart kitchens, as in use case~2, access aggregate data on the users' impairments and the list and taxonomy of smart objects, from which they select the objects suited to those impairments. The smart objects also read data from and write data to the ontology: in use case~1, the temperature sensor sends the bedroom temperature to the tablet, and the load of the cycle-ergometer is set from the user's cardiorespiratory fitness.

\paragraph{Requirements.}
Two non-functional requirements emerged from the use cases: health conditions must be represented with the ICF, and the measurements of smart objects, including those of the sensors in use case~1, must be expressed in a standard way. The functional requirements take the form of competency questions (CQs), the questions an ontology should be able to answer~\citep{gruninger1995}. Table~\ref{tab:cq} lists an excerpt grouped into three categories, with the answers used in the ontology; both the questions and the answers derive from the use cases.

\begin{table}[tbp]
\centering\small
\caption{Competency questions and their answers.}\label{tab:cq}
\begin{tabularx}{\textwidth}{@{}l>{\raggedright\arraybackslash}p{0.36\textwidth}>{\raggedright\arraybackslash}X@{}}
\toprule
& Question & Answer \\
\midrule
\multicolumn{3}{@{}l}{\emph{Group 1: users, their health condition, and their cardiorespiratory fitness}}\\
CQ1 & Who are the human end-users of the D4All system? & Simona and a caregiver (use case~1); Silvia, Sara, and Antonio (use case~2). They are elderly and impaired users.\\
CQ2 & Which users must be modeled in the ontology? & Simona (use case~1); Silvia, Sara, and Antonio (use case~2).\\
CQ3 & How are users identified in the VHF? & By an identity record for each user.\\
CQ4 & Which data form the identity record? & Name, surname, address, telephone number, email, birthplace, tax identification number.\\
CQ5 & How is a user's health condition classified? & With the ICF.\\
CQ6 & How is a user's cardiorespiratory fitness assessed? & By a specific test conducted with clinical personnel; the resulting data must be represented in the ontology.\\
\midrule
\multicolumn{3}{@{}l}{\emph{Group 2: smart objects of the house}}\\
CQ7 & Which smart objects are involved in the two use cases? & Use case~1: cycle-ergometer, environmental thermometer, air quality sensor, presence sensor, heart rate monitor, breath rate monitor, blood pressure meter, tablet. Use case~2: ovens, dishwashers, cooktops, refrigerators, presence sensors, air quality sensors, luminance sensors.\\
CQ8 & How are smart objects identified in the VHF? & By hardware version, MAC address, software version, textual description, manufacturer, and model, together with the list of the object's programs and its location in the house.\\
CQ9 & How is a smart object judged suitable for the impairments of a type of user? & Each smart object is assigned to specific categories on the basis of its features and programs.\\
CQ10 & How are the configuration projects of use case~2 managed? & For each selected smart object it must be possible to state the projects in which it is used.\\
\midrule
\multicolumn{3}{@{}l}{\emph{Group 3: measurements of smart objects}}\\
CQ11 & Which measurements arise from the two use cases? & Use case~1: temperature, pulse rate, breath rate, systolic and diastolic blood pressure, CO\textsubscript{2} concentration in the air, presence of a user in a room. Use case~2: luminance in a room.\\
CQ12 & How are the measurements described? & With standards for the description of measurements.\\
\bottomrule
\end{tabularx}
\end{table}

\paragraph{Terminology.}
The last task extracts the terms of the CQs and their answers into a pre-glossary. These terms were later formalized as concepts, instances, and relationships. Frequent terms indicate the domains for which existing, reusable knowledge sources should be sought.

\section{Development with the NeOn methodology}\label{sec:neon}

The NeOn methodology builds a network of interconnected ontologies linked by domain-dependent relationships. It is organized around nine scenarios that may occur during ontology development~\citep{suarez2015}, ranging from development from scratch to the reuse of existing models (Table~\ref{tab:neon}). The VHDM was developed under scenarios~1 and~4: some ontologies were written from scratch, and existing ones were adapted to the needs of the project. Three reusable ontologies were identified: the ICF Ontology,\footnote{\url{https://bioportal.bioontology.org/ontologies/ICF}} the Units Ontology (UO)~\citep{gkoutos2012}, and the Vital Sign Ontology (VSO)~\citep{goldfain2011}. The ontology is written in English.

\begin{table}[tbp]
\centering\small
\caption{The nine scenarios of the NeOn methodology~\citep{suarez2015}. The VHDM uses scenarios~1 and~4.}\label{tab:neon}
\begin{tabularx}{\textwidth}{@{}c>{\raggedright\arraybackslash}p{0.36\textwidth}>{\raggedright\arraybackslash}X@{}}
\toprule
& Scenario & Description \\
\midrule
\textbf{1} & \textbf{Development from scratch} & Starts from the ORSD and produces a new ontology.\\
2 & Reuse and reengineering of non-ontological resources & Taxonomies, thesauri, schemes, and other non-ontological resources of the domain are transformed into ontologies or contribute to their development.\\
3 & Reuse of ontological resources & Existing ontologies are reused as they are.\\
\textbf{4} & \textbf{Reuse and reengineering of ontological resources} & Existing ontologies are reengineered for the intended purpose.\\
5 & Reuse and merging of ontological resources & Several ontologies of the same domain are merged, mapped, or aligned.\\
6 & Reuse, merging, and reengineering of ontological resources & The result of merging is reengineered for the intended purpose.\\
7 & Reuse of ontology design patterns & Design patterns~\citep{gangemi2005,gangemi2009} are reused during conceptualization.\\
8 & Restructuring of ontological resources & Resources to be integrated are modularized, pruned, extended, or specialized.\\
9 & Localization of ontological resources & The terms of the ontology are translated into other natural languages.\\
\bottomrule
\end{tabularx}
\end{table}

The ORSD and the selected scenarios led to an architecture with two main domains, each realized as an ontology: the User's Health Condition Assessment ontology (Section~\ref{sec:health}) and the Smart Objects ontology (Section~\ref{sec:smart}).

\section{The User's Health Condition Assessment ontology}\label{sec:health}

\begin{figure}[tbp]
\centering
\includegraphics[width=0.45\textwidth]{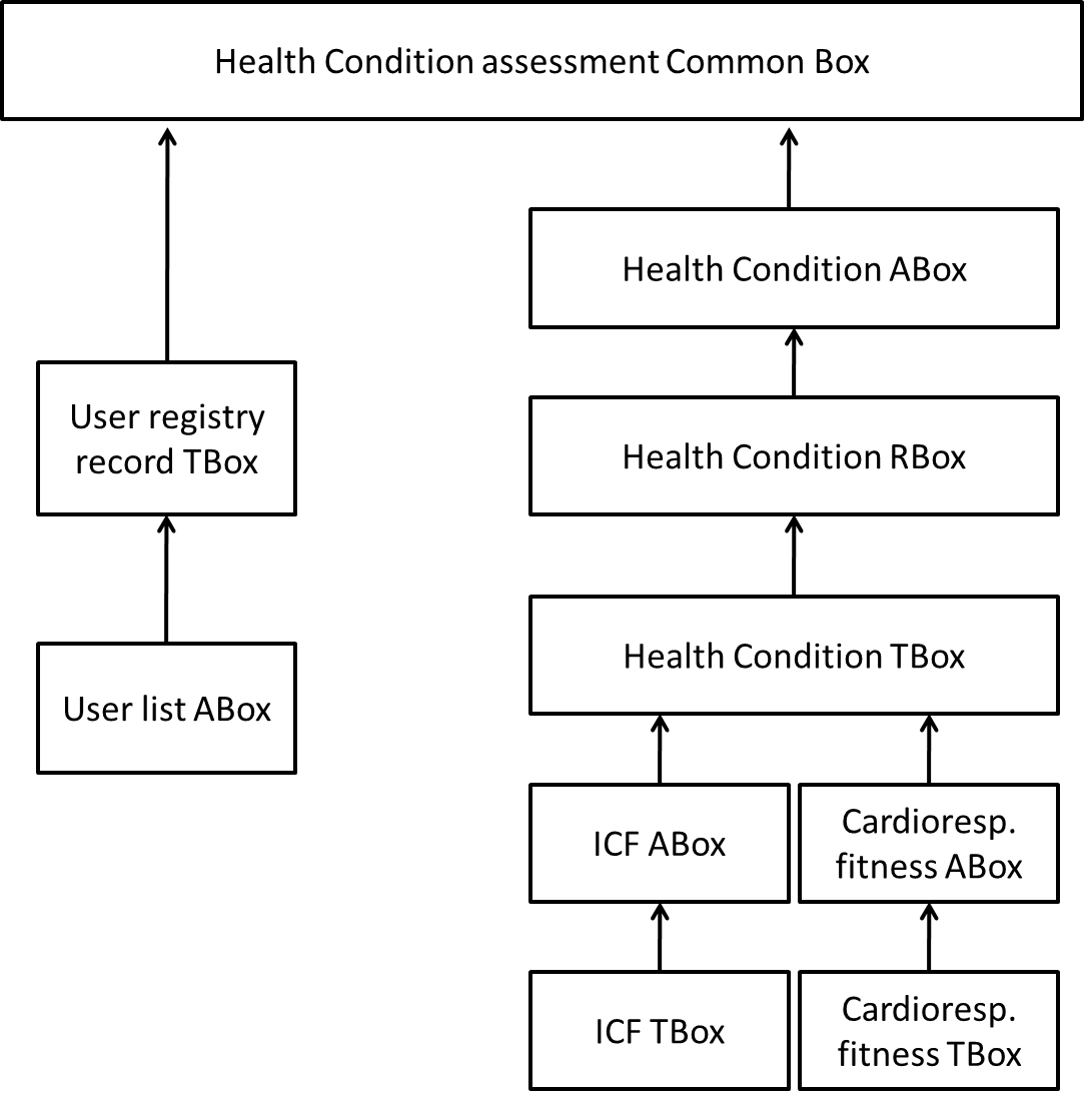}
\caption{Architecture of the User's Health Condition Assessment ontology, which gives an ICF-based description of the user and of the user's cardiorespiratory fitness.}\label{fig:health}
\end{figure}

This ontology describes the users, their health condition, and their cardiorespiratory fitness. It has two modules (Figure~\ref{fig:health}): the users' identity records and the health condition. The latter is divided into an ICF-based health condition sub-module and a cardiorespiratory fitness sub-module.

\subsection{Identity records}\label{sec:identity}

Each user of the D4All project is identified by an identity record, a kind of identity card holding static data. The TBox of this module contains the class \ont{User} with the disjoint subclasses \ont{Female} and \ont{Male}, the object property \ont{hasHealthCondition}, which links a user to a health condition, and ten datatype properties with domain \ont{User}: street, city, and country of the address, date of birth, age in years, email address, first name(s), surname, telephone number, and tax identification number. The ABox, called User List, holds the users and their identity records. \ont{User} has a further subclass \ont{ImpairedUser}, which groups \ont{VisionImpairedUser}, \ont{MotorImpairedUser}, \ont{CognitiveImpairedUser}, and \ont{HearingImpairedUser}; each of these is subdivided by grade of impairment. The object property \ont{hasCurrentLocation} records the room in which the user currently is (Section~\ref{sec:smart}).

\subsection{Health condition}\label{sec:icf}

\begin{figure}[tbp]
\centering
\includegraphics[width=0.55\textwidth]{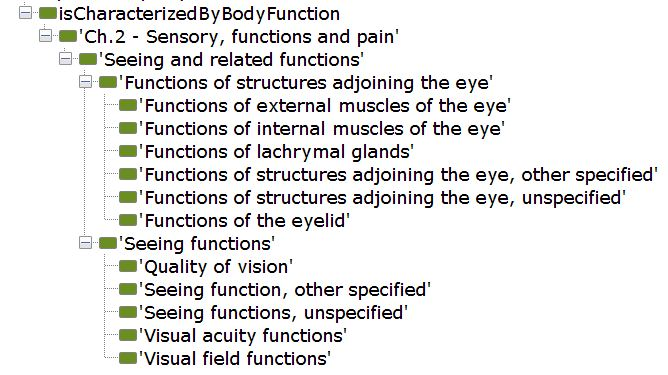}
\caption{Excerpt of the reengineered ICF ontology: datatype properties for the second chapter (b2) of the Body Functions component, down to third-level categories such as Quality of vision.}\label{fig:icfprops}
\end{figure}

The health condition of a user consists of the physiological situation assessed with the ICF and of the user's cardiorespiratory fitness. The ICF part reuses the ICF Ontology, which was heavily pruned and reengineered for the scope of the project. In the original ontology the ICF codes are individuals. In the VHDM they are datatype properties, so that each health condition can carry its own value for the same code. The result is a simplified TBox of properties for the concepts originally modeled under the class \ont{ICFCategory} and its subclasses. There are datatype properties for the four components of the ICF (Body Functions, Activities and Participation, Environmental Factors, Body Structures); below each component are its chapters, and below each chapter are sub-properties for the second- and third-level categories (Figure~\ref{fig:icfprops}). Each property links a health condition to an integer in $\{0,1,2,3,4\}$, the generic ICF qualifier of the severity of an impairment, where 0 means no impairment and 4 complete impairment. For example, the health condition of the user in Figure~\ref{fig:user} has the value~2 (moderate impairment) for Visual acuity functions and the value~1 (mild impairment) for Functions of the eyelid.

\begin{figure}[tbp]
\centering
\includegraphics[width=0.58\textwidth]{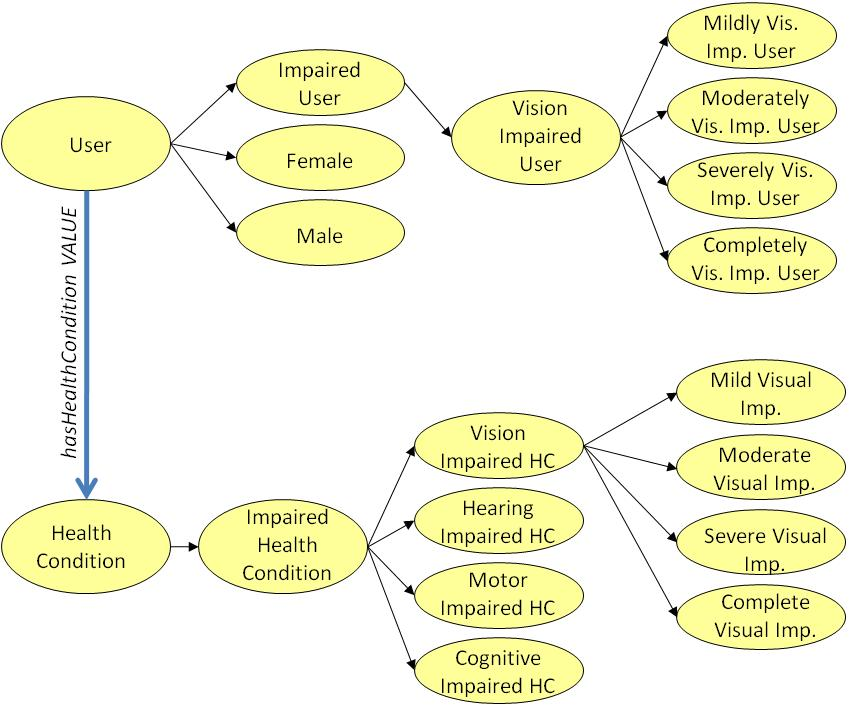}
\caption{Excerpt of the TBox: the subclasses of \ont{VisionImpairedUser} and the classification of \ont{HealthCondition}.}\label{fig:hcclasses}
\end{figure}

In the TBox, \ont{HealthCondition} has the subclass \ont{ImpairedHealthCondition}, which is divided into \ont{VisionImpairedHealthCondition}, \ont{MotorImpairedHealthCondition}, \ont{HearingImpairedHealthCondition}, and \ont{CognitiveImpairedHealthCondition}. Each of these is subdivided by grade of impairment according to the ICF qualifiers (Figure~\ref{fig:hcclasses}). Since a user is linked to a health condition through \ont{hasHealthCondition}, a reasoner can infer whether the user is impaired and with which grade.

\subsection{Cardiorespiratory fitness}\label{sec:cf}

The CF module, composed of a TBox, an ABox, and an RBox, links each \ont{User} individual to an individual representing the user's cardiorespiratory fitness (CF) and classifies the latter automatically. CF is the ability to perform exercise for prolonged periods~\citep{acsm2013}. It is measured as \vo, the maximum rate of oxygen consumption by the muscles involved in the exercise~\citep{glass2007}. Table~\ref{tab:vo2} gives the reference percentiles of \vo{} by sex and age, and Table~\ref{tab:cfclass} maps the percentile of a user's \vo{} to a CF class.

\begin{table}[tbp]
\small
\begin{minipage}[t]{0.56\textwidth}
\centering
\caption{Reference percentiles of \vo{} (ml/min/kg) by sex and age (adapted from~\citealp{glass2007}).}\label{tab:vo2}
\begin{tabular}{@{}llccccc@{}}
\toprule
& & \multicolumn{5}{c}{Percentile}\\
\cmidrule(l){3-7}
Sex & Age (years) & 20th & 40th & 60th & 80th & 95th\\
\midrule
\multirow{6}{*}{Men}
& 20--29 & 38.1 & 42.2 & 45.7 & 51.1 & 56.2\\
& 30--39 & 36.7 & 41.0 & 44.4 & 47.5 & 54.3\\
& 40--49 & \chk{34.6} & 38.4 & 42.4 & 46.8 & 52.9\\ 
& 50--59 & 31.1 & 35.2 & 38.3 & 43.3 & 49.7\\
& 60--69 & 27.4 & 31.4 & 35.0 & 39.5 & 46.1\\
& 70--79 & 23.7 & 28.0 & 30.9 & 36.0 & 42.4\\
\midrule
\multirow{6}{*}{Women}
& 20--29 & 31.6 & 35.5 & 39.5 & 44.0 & 50.2\\
& 30--39 & 29.9 & 33.8 & 35.7 & 41.0 & 46.9\\
& 40--49 & 28.0 & 31.6 & 35.1 & 38.9 & 45.2\\
& 50--59 & 25.5 & 28.7 & 31.4 & 35.2 & 39.9\\
& 60--69 & 23.7 & 26.6 & 29.1 & 32.3 & 36.9\\
& 70--79 & 21.2 & 23.8 & 26.6 & 30.2 & 36.7\\
\bottomrule
\end{tabular}
\end{minipage}\hfill
\begin{minipage}[t]{0.4\textwidth}
\centering
\caption{Cardiorespiratory fitness classes as a function of the percentile $p$ of the user's \vo{} in Table~\ref{tab:vo2}.}\label{tab:cfclass}
\begin{tabular}{@{}ll@{}}
\toprule
CF class & Percentile $p$\\
\midrule
Seriously limited & $p \le 20$\\
Reduced & $20 < p \le 40$\\
Adequate & $40 < p \le 60$\\
Good & $60 < p \le 80$\\
Excellent & $80 < p \le 95$\\
More than excellent & $p > 95$\\
\bottomrule
\end{tabular}
\end{minipage}
\end{table}

A short test on the cycle-ergometer, supervised by clinical personnel, gives the user's \vo{} and maximum acceptable work load. A set of SWRL rules then classifies the \vo{} value according to Table~\ref{tab:cfclass}. For instance, rule~\rref{r:slcf} states that a woman aged between 70 and 79 whose \vo{} is below 21.2\,ml/min/kg, the 20th percentile of her group, has a CF in the class \ont{SeriouslyLimitedCF}:

\begin{swrl}{r:slcf}
Female(?p), isCharacterizedByCF(?p,~?cf), hasAge(?p,~?age), greaterThanOrEqual(?age,~70), lessThanOrEqual(?age,~79), hasVO2max(?cf,~?vo2), lessThan(?vo2,~21.2)\imp SeriouslyLimitedCF(?cf)
\end{swrl}

\noindent Further SWRL rules derive the exercise load of the cycle-ergometer from the CF class. Rule~\rref{r:load} covers a woman with an impaired health condition and a seriously limited CF; with $w$ her maximum work load, the built-in \texttt{multiply} gives $v = 30w$ and \texttt{divide} gives the suggested load $v/100 = 0.3\,w$:

\begin{swrl}{r:load}
Female(?p), hasHealthCondition(?p,~?hc), ImpairedHealthCondition(?hc), isCharacterizedByCF(?p,~?cf), SeriouslyLimitedCF(?cf), hasMaxWorkLoad(?cf,~?w), divide(?load,~?v,~100), multiply(?v,~?w,~30)\imp hasSuggestedCycle-ergometerLoad(?cf,~?load)
\end{swrl}

\noindent The suggested load is 30\% of the estimated maximum work load for an impaired user and 50\% for a user without impairments. The exercise lasts 15 minutes for impaired users and 25 minutes for users without impairments.

\begin{figure}[tbp]
\centering
\includegraphics[width=0.8\textwidth]{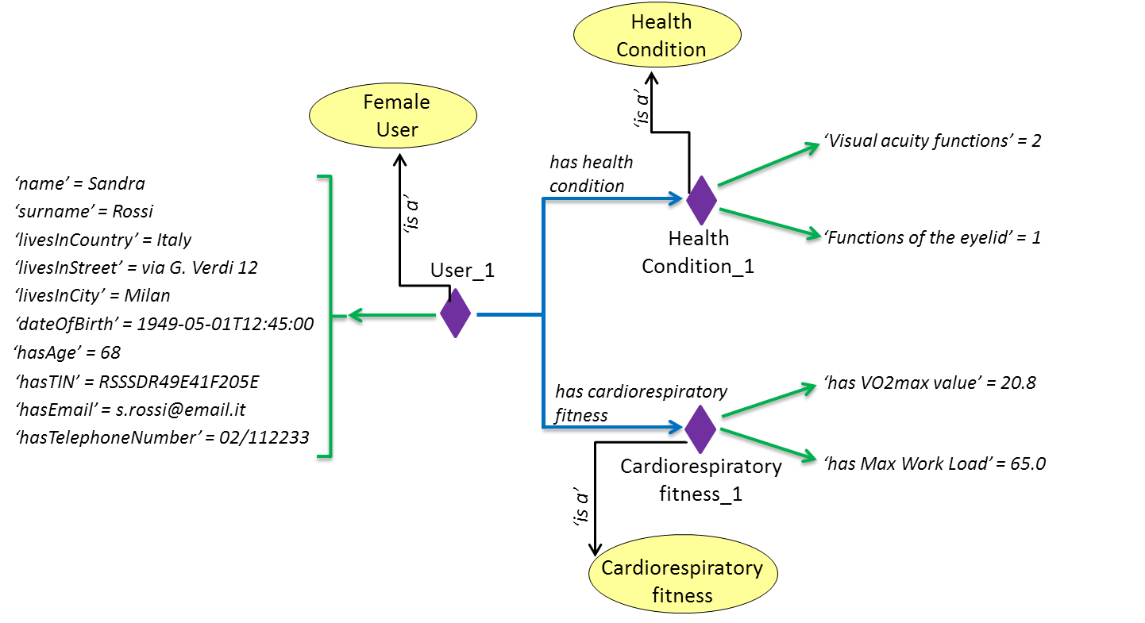}
\caption{The individual \ont{User\_1} with her identity record, health condition, and cardiorespiratory fitness.}\label{fig:user}
\end{figure}

The health condition and the CF together define the health status of the user: the impairments and the aerobic fitness. Figure~\ref{fig:user} shows a complete \ont{User} individual, a woman aged 68 with a \vo{} of 20.8\,ml/min/kg and a maximum work load of 65.0. Her \vo{} is below 23.7, the 20th percentile of women aged 60--69 in Table~\ref{tab:vo2}, so the rule of her age group, analogous to~\rref{r:slcf}, classifies her CF as \ont{SeriouslyLimitedCF}. Her health condition, with qualifier~2 for Visual acuity functions, is impaired, and rule~\rref{r:load} therefore suggests a load of $0.3 \times 65.0 = 19.5$. With the Pellet reasoner\footnote{\url{https://www.w3.org/2001/sw/wiki/Pellet}} it is possible to infer whether a user has an impairment, its grade, and the user's CF class. The ICF qualifiers are assigned by medical personnel, and \vo{} is obtained from a physical test supervised by clinicians.

\section{The Smart Objects ontology}\label{sec:smart}

\begin{figure}[tbp]
\centering
\includegraphics[width=0.55\textwidth]{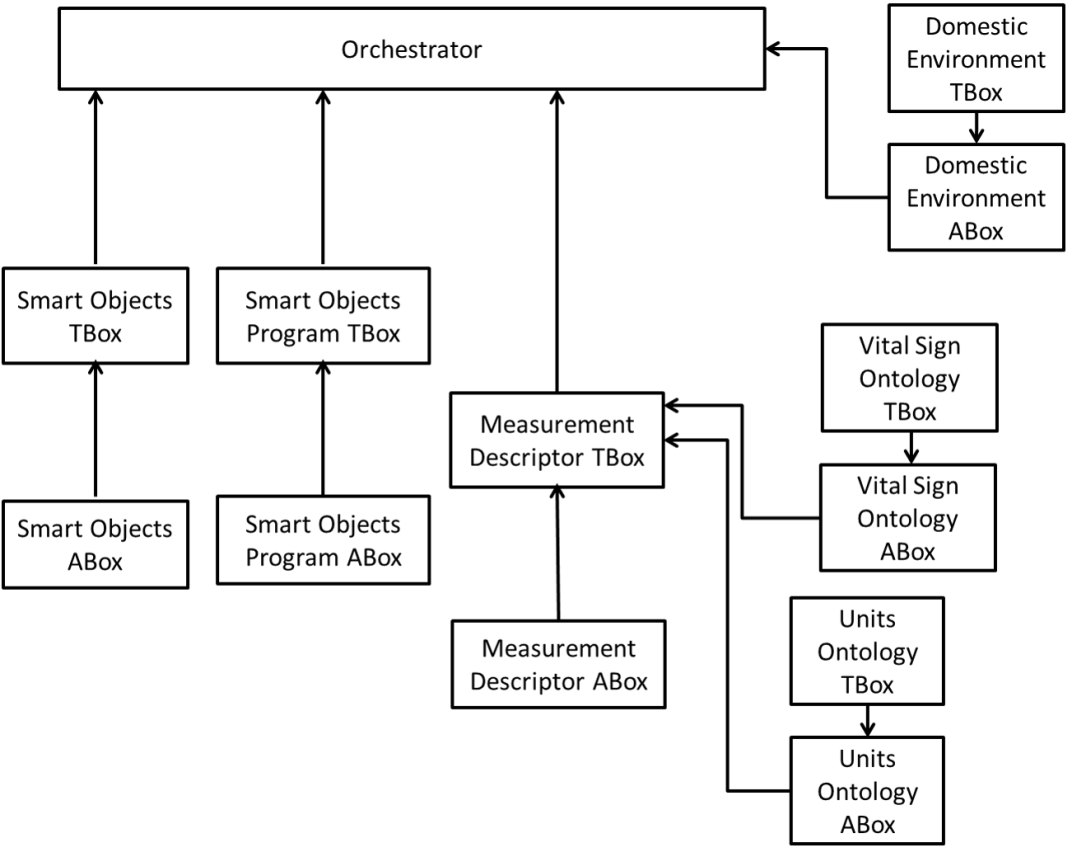}
\caption{Architecture of the Orchestrator ontology: descriptions of the smart objects and of their programs, of the domestic environment, and of the measurements performed by the smart objects.}\label{fig:orch}
\end{figure}

The second ontology, called Orchestrator, describes the smart objects, links them to the rooms of the house, and lists the measurements they perform. It has three main modules (Figure~\ref{fig:orch}): the Domestic Environment module, which lists and describes the rooms of the user's home; the Smart Object module, which lists and describes the smart objects and their programs; and the Measurement Descriptor module, which formalizes the measurements performed by each smart object. A further module, imported into the Orchestrator, supports the configuration of living environments.

\subsection{Smart objects and their programs}\label{sec:solist}

The Smart Object module consists of a TBox and an ABox that collect the smart objects in the user's home. Datatype properties record, for each object, the application and physical protocols, hardware and software versions, MAC address, manufacturer, model, and status (on or off). Each object is classified by type: \ont{Cooktop}, \ont{Dishwasher}, \ont{Oven}, \ont{Refrigerator}, or \ont{Sensor}, the last with the subclasses \ont{AirQualitySensor}, \ont{LuminanceSensor}, \ont{PresenceSensor}, and \ont{ThermohygrometricSensor} (Figure~\ref{fig:soclasses}).

The object property \ont{hasProgram} links a smart object to its programs, and its inverse \ont{isProgramOf} links a program to the objects that run it; a program can belong to several smart objects of the same type. The programs have the subclasses \ont{DishwasherProgram} and \ont{OvenProgram}, which reflect the needs identified by the project partners for the kitchen use cases. Datatype properties give each program a name and a description and set its starting time or the temperature the smart object must keep while the program runs. For example, the individual \ont{Static} of class \ont{OvenProgram} sets the oven temperature to 250.0\,\textdegree C.

\subsection{Measurements}\label{sec:measurements}

Each smart object produces one or more measurements, which may concern the domestic environment, a smart object, or a user. A thermometer in the living room, for instance, measures the temperature of the room, whereas a thermometer worn by the user measures body temperature. The Measurement Descriptor module therefore divides measurements into three subclasses: \ont{EnvironmentalMeasurement} for the domestic environment, \ont{VitalSign} for the user's physiological status, and \ont{SmartObjectMeasurement} for the status of a smart object. Units are described with the Units Ontology~\citep{gkoutos2012}, extended with a TBox of the physical magnitudes that each unit measures, linked through the object property \ont{isUnitOf}. The module is imported into a simplified version of the Vital Sign Ontology~\citep{goldfain2011} that keeps the four vital signs relevant to the project: \ont{BloodPressure}, with subclasses \ont{DiastolicBloodPressure} and \ont{SystolicBloodPressure}; \ont{PulseRate}; \ont{RespiratoryRate}; and \ont{BodyTemperature}, with subclasses \ont{AxillaryTemperature}, \ont{OralTemperature}, \ont{RectalTemperature}, and \ont{TympanicTemperature}.

\begin{figure}[tbp]
\centering
\includegraphics[width=0.62\textwidth]{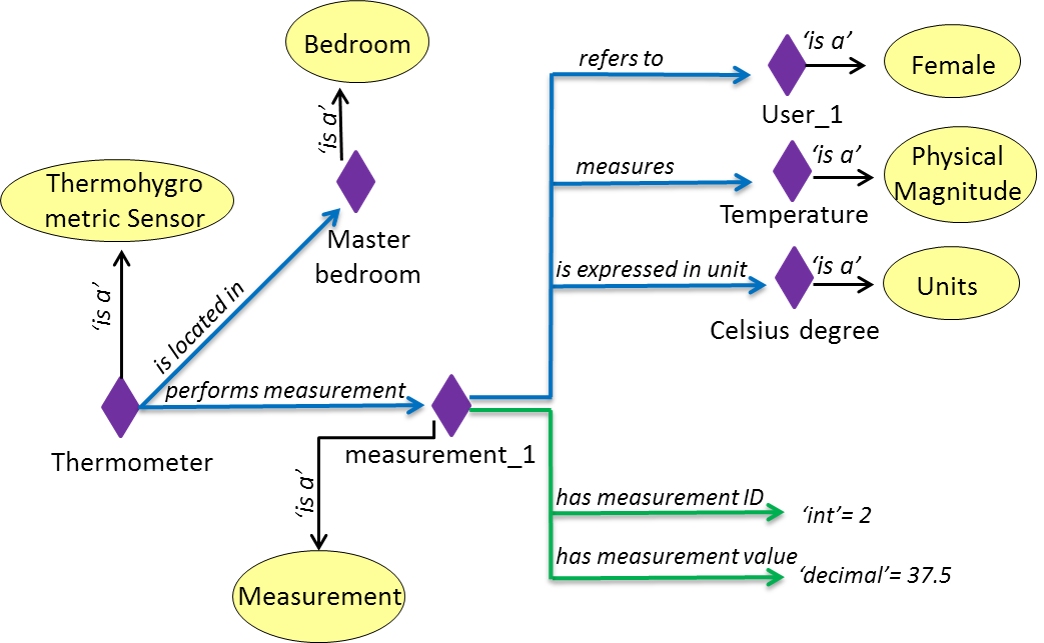}
\caption{A thermometer individual performing a measurement that refers to a user.}\label{fig:meas}
\end{figure}

Each measurement is described by object and datatype properties. The object property \ont{refersTo} indicates whether the measurement concerns a person, a room, or a smart object; \ont{measures} gives the physical magnitude, and \ont{isExpressedInUnit} the unit. The datatype properties \ont{hasMeasurementID}, \ont{hasMeasurementValue} (with range \texttt{xsd:decimal}), \ont{hasDescription}, and \ont{hasAssessmentDate} give the identifier, value, description, and date and time of the measurement. SWRL rules classify a measurement as an \ont{EnvironmentalMeasurement}, a \ont{VitalSign}, or a \ont{SmartObjectMeasurement} according to whether the individual it refers to belongs to \ont{Room}, \ont{User}, or \ont{SmartObjects}. In Figure~\ref{fig:meas}, a thermometer located in the master bedroom performs \ont{measurement\_1}, which refers to \ont{User\_1}, measures the magnitude \ont{Temperature} in Celsius degrees, and has the value 37.5; since it refers to a user, the rules classify it as a \ont{VitalSign}. Because of this classification, a smart object that needs a certain kind of measurement can search the individuals of the corresponding class.

\subsection{Comfort dimensions}\label{sec:comfort}

The description of measurements also makes it possible to describe the comfort of the home. The class \ont{ComfortDimension} has the subclasses \ont{CO2Concentration}, \ont{DomesticHumidityRate}, \ont{DomesticTemperature}, and \ont{DomesticLuminance}; each one classifies environmental measurements against limits taken from standards.

\paragraph{CO\textsubscript{2} concentration.}
The class \ont{CO2Concentration} has two subclasses. The first, \ont{AcceptableCO2Concentration}, collects the individuals of \ont{EnvironmentalMeasurement} that measure the physical magnitude \ont{ConcentrationOf} in parts per million (ppm) with a value below 1000. The second, \ont{UnacceptableCO2Concentration}, is its complement and collects the same kind of measurements with a value of 1000\,ppm or more. These ranges summarize the limits of the Italian standard UNI EN 13779~\citep{uni2008}.

\paragraph{Humidity.}
\ont{DomesticHumidityRate} has two subclasses for the acceptable relative humidity in a room in summer and in winter. The seasons are modeled as individuals with the datatype properties \ont{hasStartTime} and \ont{hasFinishTime}. Since winter spans the turn of the year, the date of a measurement is compared with three periods delimited by the two seasons, and its value with the limits of the Italian regulations, which correspond to ASHRAE Standard 55-2010~\citep{ashrae2010}. Rules~\rref{r:hsum} and~\rref{r:hwin} accept a relative humidity between 20\% and 60\% in summer and between 25\% and 60\% in the winter interval represented by \ont{Winter\_season\_B}:

\begin{swrl}{r:hsum}
EnvironmentalMeasurement(?m), measures(?m,~PhMag\_Relative\_humidity), hasStartTime(Summer\_season,~?start), hasFinishTime(Summer\_season,~?end), hasAssessmentDate(?m,~?t), greaterThanOrEqual(?t,~?start), lessThanOrEqual(?t,~?end), hasMeasurementValue(?m,~?v), greaterThanOrEqual(?v,~20), lessThanOrEqual(?v,~60)\imp SummerAcceptableRateHumidity(?m)
\end{swrl}
\begin{swrl}{r:hwin}
EnvironmentalMeasurement(?m), measures(?m,~PhMag\_Relative\_humidity), hasStartTime(Winter\_season\_B,~?start), hasFinishTime(Winter\_season\_B,~?end), hasAssessmentDate(?m,~?t), greaterThanOrEqual(?t,~?start), lessThanOrEqual(?t,~?end), hasMeasurementValue(?m,~?v), greaterThanOrEqual(?v,~25), lessThanOrEqual(?v,~60)\imp WinterAcceptableRateHumidity(?m)
\end{swrl}

\paragraph{Temperature.}
The class \ont{DomesticTemperature} is divided into \ont{UncomfortableTemperature} and \ont{ComfortableTemperature}; the latter has the seasonal subclasses \ont{SummerComfortableTemperature} and \ont{WinterComfortableTemperature}. In winter, comfort depends on whether domestic heating is on, so an individual \ont{Non\_Heating\_Period} represents the period of the year in which heating may not be turned on in Italy, with datatype properties for its start and finish (for example \mbox{2015-04-15T00:00:01} and \mbox{2015-10-14T23:59:59}). The limits of the Italian regulations, which correspond to those of ASHRAE Standard 55-2010~\citep{ashrae2010}, are encoded in rules~\rref{r:tsum}--\rref{r:twin2}: a temperature between 23 and 27\,\textdegree C is comfortable within the non-heating period, and one between 20 and 24\,\textdegree C is comfortable after or before it.

\begin{swrl}{r:tsum}
EnvironmentalMeasurement(?m), hasAssessmentDate(?m,~?t), hasStartTime(Non\_Heating\_Period,~?start), hasFinishTime(Non\_Heating\_Period,~?end), greaterThan(?t,~?start), lessThan(?t,~?end), hasMeasurementValue(?m,~?v), greaterThanOrEqual(?v,~23), lessThanOrEqual(?v,~27)\imp SummerComfortableTemperature(?m)
\end{swrl}
\begin{swrl}{r:twin1}
EnvironmentalMeasurement(?m), hasAssessmentDate(?m,~?t), hasStartTime(Non\_Heating\_Period,~?start), hasFinishTime(Non\_Heating\_Period,~?end), greaterThan(?t,~?end), hasMeasurementValue(?m,~?v), greaterThanOrEqual(?v,~20), lessThanOrEqual(?v,~24)\imp WinterComfortableTemperature(?m)
\end{swrl}
\begin{swrl}{r:twin2}
EnvironmentalMeasurement(?m), hasAssessmentDate(?m,~?t), hasStartTime(Non\_Heating\_Period,~?start), hasFinishTime(Non\_Heating\_Period,~?end), lessThan(?t,~?start), hasMeasurementValue(?m,~?v), greaterThanOrEqual(?v,~20), lessThanOrEqual(?v,~24)\imp WinterComfortableTemperature(?m)
\end{swrl}

\paragraph{Luminance.}
UNI EN 12464-1~\citep{uni2004} recommends illuminance levels that depend on the task performed in a room. A task that requires precision and attention needs at least 500\,lux on the working area, such as the cooktop in the kitchen or the desk in the studio; other tasks need at least 100\,lux. \ont{DomesticLuminance} collects measurements in lux of the luminance of a room and has two subclasses. \ont{ComfortableLuminanceForWorkingAreas} contains the measurements of at least 500\,lux that refer to the rooms in which users perform demanding activities of daily living: bathroom, dining room, kitchen, studio, and living room. \ont{ComfortableLuminanceForNon-workingAreas} contains the measurements of at least 100\,lux that refer to the other rooms: basement, bedroom, entrance, hallway, storage closet, attic, and balcony.

\medskip
Classifying environmental measurements under \ont{ComfortDimension} keeps track of every air quality and temperature measurement in a room. In use case~1, this allows the user to exercise on the cycle-ergometer in a safe environment. The environmental measurements produced by sensors and smart objects are also stored for later analysis of comfort.

\subsection{Configuration of living environments}\label{sec:config}

\begin{figure}[tbp]
\centering
\includegraphics[width=0.62\textwidth]{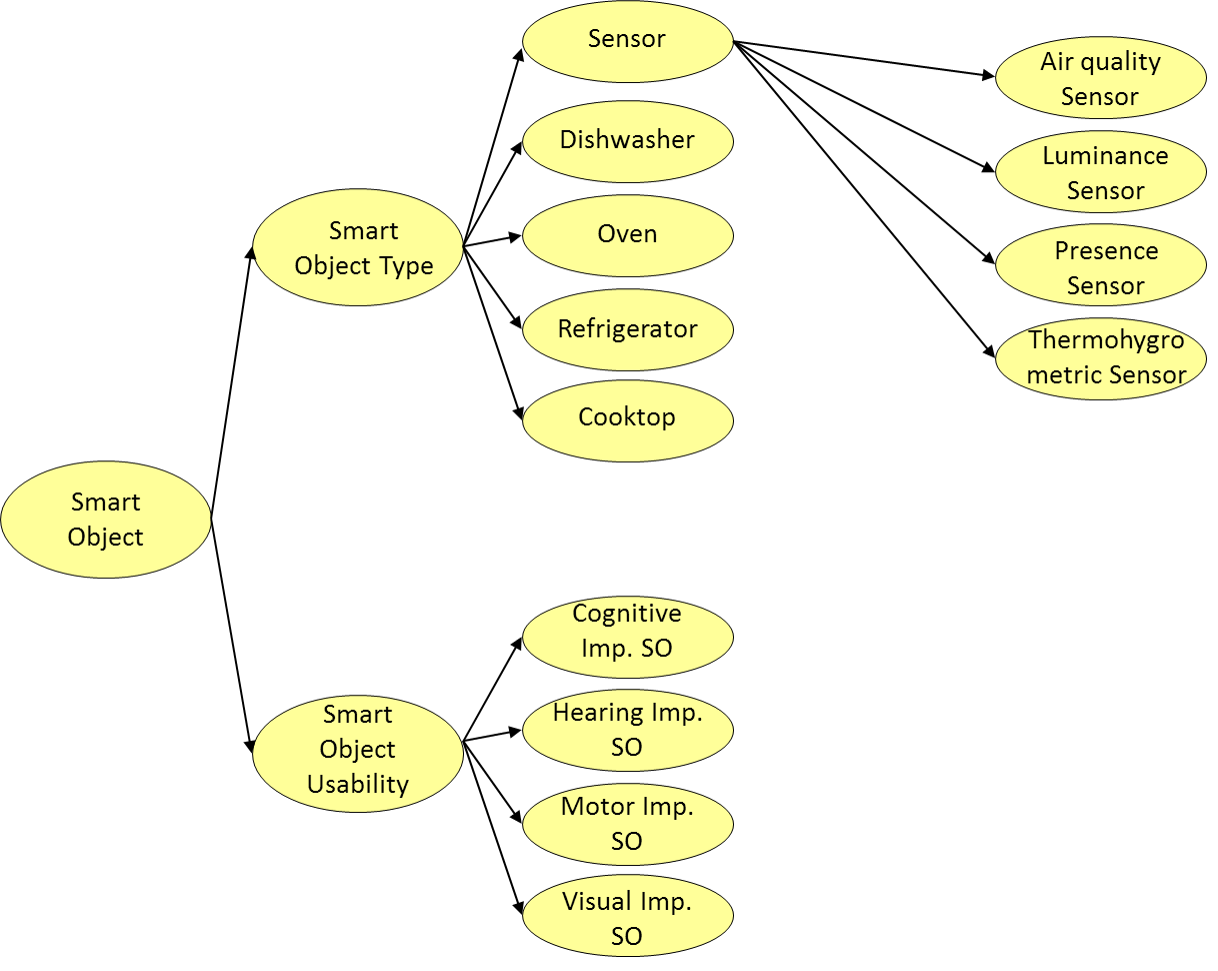}
\caption{Excerpt of the TBox: smart objects classified by type and by suitability for specific categories of impaired users.}\label{fig:soclasses}
\end{figure}

The configuration module classifies the domestic smart objects by their suitability for impaired users. The project partners identified the smart objects whose design and features help users cope with their impairments in activities of daily living. Smart objects are therefore classified both by type (Section~\ref{sec:solist}) and by suitability for specific categories of impaired users (Figure~\ref{fig:soclasses}).

\begin{figure}[tbp]
\centering
\includegraphics[width=0.65\textwidth]{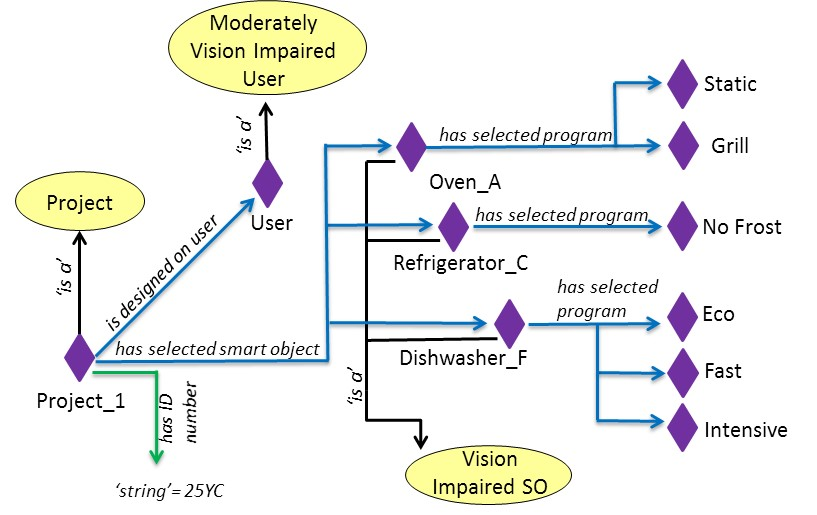}
\caption{A \ont{Project} individual for a moderately vision-impaired user, with the selected smart objects and programs.}\label{fig:project}
\end{figure}

The module also describes the projects that configure a living environment, as in use case~2. Each project is an individual with an identifier (\ont{hasIDNumber}); the object properties \ont{hasSelectedSmartObject} and \ont{hasSelectedProgram} record the chosen smart objects and programs, and \ont{isDesignedOnUser} records the user for whom the project is designed. Figure~\ref{fig:project} shows \ont{Project\_1}, designed for a moderately vision-impaired user. It selects three kitchen appliances suitable for vision-impaired users, \ont{Oven\_A}, \ont{Refrigerator\_C}, and \ont{Dishwasher\_F}, together with their programs (for the oven, \ont{Static} and \ont{Grill}). Imported into the Orchestrator, this module lets SWRL rules infer which smart objects suit a given category of impaired users. The designer then chooses from the inferred list, and the reasoning results act as decision support for the design of the living environment.

\section{Conclusions and future work}\label{sec:conclusions}

We have described the development of the ontology of the Design For All project. The process starts from use cases in ambient assisted living, which identify the basic concepts the ontology must contain. The NeOn methodology and the ORSD then refine these concepts and produce a list of competency questions. The resulting modeling choices support both use cases with concepts and relations that describe the activities they portray.

The ontology centers on the user's health profile, expressed with an international standard (the ICF) and complemented by an important physiological indicator, cardiorespiratory fitness. It also describes the smart objects of a house and their measurements. On this basis it models domestic comfort metrics and serves as a configuration tool with which designers offer each category of users the most suitable smart objects. The applications and devices of the Design For All project query the data through SPARQL.

Future work will extend the use of use cases in ontology engineering and add use cases with new types of users, such as users without impairments and users with mixed impairments, new smart objects, and other activities in the home supported by the Virtual Home Framework. Two further developments are the customization of comfort metrics according to the preferences of each user and an extension of the cardiorespiratory fitness model toward customized and dynamic work loads. The modular architecture of the ontology is expected to ease both.

\FloatBarrier
\bibliographystyle{plainnat}
\bibliography{refs}

\end{document}